\documentclass[journal,twoside,10pt]{IEEEtran}
\usepackage{amsmath,amssymb}
\usepackage[T1]{fontenc}
\usepackage[dvips]{graphicx, color}
\usepackage{subfigure}
\usepackage{latexsym}
\definecolor{Bgreen}{rgb}{ .0, .55, .0 }
\definecolor{Red}{rgb}{ 1.0, .0, .0 }
\definecolor{Navy}{rgb}{ 0.0, .0, 1.0 }

\newcommand{\eqa}{\begin{eqnarray}}
\newcommand{\ena}{\end{eqnarray}}

\def\rn#1{\expandafter{\romannumeral#1}} 

\def\Cc{Chua's circuit}
\def\CDS{chaotic dynamical system}
\newcounter{linenumber}
    {\end{list}}

\begin{document}
\pagestyle{empty}




%
\title{Compressive sampling with chaotic dynamical systems}
%
%
%

\author{Venceslav Kafedziski,~\IEEEmembership{Member,~IEEE}~and~
Toni~Stojanovski,~\IEEEmembership{Member,~IEEE}
\thanks{V.\ Kafedziski is with the Faculty of Electrical Engineering and Information Technologies, University Ss Cyril and Methodius, 1000 Skopje, Republic of Macedonia. (phone: +389 2 3099120, e-mail kafedzi@feit.ukim.edu.mk)}
\thanks{T.\ Stojanovski is with the Faculty of Informatics, European University, bld Kliment Ohridski 68, 1000 Skopje, Republic of Macedonia. (phone: +389 78 396 693, e mail: toni.stojanovski@eurm.edu.mk)}
\vspace{-.5cm} }
\maketitle \thispagestyle{empty} 
\begin{abstract}
We investigate the possibility of using different chaotic sequences to construct  measurement matrices in compressive sampling.
In particular, we consider sequences generated by Chua, Lorenz and R\"ossler dynamical systems and investigate the accuracy of reconstruction when using 
each of them to construct measurement matrices. 
Chua and Lorenz sequences appear to be suitable to construct measurement matrices. 
We compare the recovery rate of the original sequence with that
obtained by using Gaussian, Bernoulli and uniformly distributed random measurement matrices. 
We also investigate the impact of correlation on the recovery rate.
It appears that correlation does not influence the probability of exact reconstruction
 significantly.
\end{abstract}


\IEEEpeerreviewmaketitle

\vspace{-.2cm}
\section{Introduction}\label{intro}
\IEEEPARstart{A}{ccording} to the Nyquist-Shannon sampling theorem,
if a signal is band limited to a bandwidth $B$, then it is completely determined by sampling it at discrete times, provided that the sampling rate is at least equal to $2B$. The original continuous-time signal can be reconstructed from the discrete-time samples via an interpolation process achieved with a low-pass filter.

Recently, work by Candes \cite{cand_feb06}, 
Donoho \cite{dono06}, and others demonstrated that it is possible to exactly reconstruct some signals from undersampled data. If a signal is sparse in the original domain, or some transform domain (sparsifying domain), meaning that it does not have many features, we can take far fewer measurements and utilize knowledge of the structure of the signal to infer the rest. Thus, we can exactly reconstruct sparse signals with far fewer measurements than needed by Nyquist - Shannon theory.

This approach to sub-Nyquist sampling has been called compressive sampling (CS). Nice overviews of compressive sampling
can be found in \cite{bara07}, \cite{cand08}. 

The idea of CS is to combine the two stages, sampling and compression. Measurements of the signal are taken using the measurement matrix, which is supposed to 
be incoherent with the matrix describing the sparsifying transform. 
More formally, the so called Restricted Isometry Property (RIP) should be met.
Since random signals are incoherent with almost anything, taking iid samples from a random distribution (for example, Gaussian or uniform) to create the measurement matrix, violates the RIP property with exponentially small probability.

Since chaotic signals exhibit similar properties to random signals, some authors have
attempted using chaotic signals to construct the measurement matrix.

In \cite{4783326} the authors continue on the work in \cite{1660793} on random filters in CS, and examine the use of chaos filters in CS with filter taps calculated from the Logistic map.  
The authors claim that their numerical simulations indicate that chaos filters generated by the Logistic map outperform random filters.

In \cite{YuBarbot-2010} the authors construct the measurement matrix with chaotic sequence from Logistic map 
and prove that, with overwhelming probability, 
the RIP of this kind of matrix is satisfied, which guarantees exact recovery. 
The authors experimentally show that chaotic matrix has similar performance to the Gaussian random matrix and sparse random matrix.
In the subsequent work \cite{5580428}, the authors show that Toeplitz-structured measurement matrix constructed using a chaotic sequence is sufficient to satisfy RIP with high probability. 
This measurement matrix can be easily built as a filter with a small number of taps.

To the best of our knowledge in the literature on using chaotic sequences in CS, 
chaotic maps other than the Logistic map have not been used. Also, in the literature 
it is usually assumed 
that the chaotic sequence should be uncorrelated, and, therefore, every $d$-th sample
is used to create the measurement matrix, where $d$ is such to ensure that the 
samples are uncorrelated. Although the performance of Logistic map has been shown
to be very satisfactory and sometimes to outperform the performance using random matrices, 
such as Gaussian, Bernoulli or uniform, it is still desirable to examine and compare
the performance of other chaotic maps.  

Therefore, in this paper we address the following questions:

- Which chaotic signals can be used to construct measuring matrices?

- Does the correlation influence the probability of exact reconstruction significantly?

- How does the performance of chaos-based measurement matrices compare to the performance of random measurement matrices?

In Section~\ref{sec:CS} we give a brief overview of compressive sampling.
Section~\ref{sec:Chaos} gives a brief overview of nonlinear dynamical systems that exhibit chaotic behaviour, 
and depicts the properties of three chaotic systems whose applicability in CS is examined.
The main results of this work are presented in Section~\ref{sec:MainResults}.
Section~\ref{sect:conclusion} concludes the paper.

\section{Compressive sampling} \label{sec:CS}
Compressive sampling answers the question if we can compress the signal ${\bf x}\in{\bf R}^N$ into some compressed basis $\Psi$ where it can be represented sparsely as a signal ${\bf s}$, then can we recover the original signal if the number of measurements $M$ is approximately equal to the number of significant components of {\bf s}? If our sparse signal ${\bf s}$ is $k$-sparse, meaning it has $k$ significant components, we can fix our solution for ${\bf s}$ in $k$ dimensions, and do some kind of optimization for the remaining $N-k$ elements. 
Mathematically, the measured samples are given by
\begin{equation}
 {\bf y} = \Phi{\bf x} = \Phi \Psi {\bf s} = \Theta {\bf s}
\end{equation}
where $\Phi$ is the measurement (sensing) basis, 
$\Psi$ is the compression basis, and $\Theta = \Phi \Psi$ is the compressive sensing matrix and is the product of the compression and measurement bases. So, we can take some small number of samples ${\bf y}$, compute the sparse representation ${\bf s}$ of our exact signal ${\bf x}$, and then apply the inverse compression approximation to recover ${\bf x}$.



Important property that $\Phi$ and $\Psi$ should meet is incoherence. The coherence measures the largest correlation between any two elements of $\Phi$ and $\Psi$. If $\Phi$ and $\Psi$ contain correlated elements, the coherence is large. Otherwise, it is small.  Mathematically, coherence is defined as,
\begin{equation}
  \mu(\Psi, \Phi)=\sqrt{N}\max _{1\leq k,j \leq N}|\langle \psi _k ,\phi _j\rangle|
\end{equation}
This function takes on values between 1 and $\sqrt{N}$.

To guarantee that the compressive sensing matrix $\Theta$ is stable, it must meet the Restricted Isometry Property (RIP):
\begin{equation}
  (1-\delta _k)\|{\bf s}\|_{l_2}^2\leq \|\Theta {\bf s}\|_{l_2}^2\leq (1+\delta _k)\|{\bf s}\|_{l_2}^2
\end{equation}
In other words, $\Theta$ must be a distance preserving transformation for all $k$-sparse vectors {\bf x}, bounded by some constant $\delta _k$, known as the restricted isometry constant. 
When this property holds, all $k$-subsets of the columns of $\Theta$ are nearly orthogonal. If this property does not hold then it is possible for a $k$-sparse signal to be in the null space of $\Theta$ and in this case it may be impossible to reconstruct these vectors.

It has been shown that bases $\Phi$ and $\Psi$ which are incoherent will satisfy RIP.


Regarding the reconstruction algorithms, different norms can be used. The $l_2$ norm does
not favor a sparse solution and has been shown that cannot be used. 
The $l_0$ norm counts the number of zeros in the vector, and that's exactly what we want to minimize. But, it turns out that we would have to try every  combination of zeros to find the solution, which is NP-hard, and thus intractable. The researchers that worked in the CS area discovered that we can solve the problem using the $l_1$  norm and obtain exact results. 

\section{Chaotic dynamical systems} \label{sec:Chaos}
Nonlinear dynamical systems are capable of exhibiting chaotic behavior for certain parameter values.
Chaotic behavior exhibits exponential sensitivity to small changes in initial conditions: two chaotic trajectories starting from arbitrarily close initial conditions will eventually diverge from each other.
Thus, even the smallest error in the measurement of initial conditions of a chaotic dynamical system precludes us from predicting its long term behavior.
Despite their deterministic definition e.g. via ordinary differential equations, chaotic dynamical systems exhibit unpredictable behaviour.
This duality in the nature of chaotic dynamical systems have sparked an immense interest in their potential applicability in a wide range of areas: cryptography, telecommunications, traffic modelling, medicine etc.
Using \CDS s for random number generation is a well researched area \cite{915385}.
In this section we present the properties of three well-known \CDS s which are relevant for generation of elements of measurement matrices in CS.
 
\subsection{Chua} \label{sec:Chua}
Properties of chaotic behaviour will be illustrated via an example
based on \Cc\, which is the simplest electronic circuit capable of exhibiting chaos.
Besides its simplicity, it is particularly useful and interesting because its
chaotic behaviour has been proven analytically, numerically and experimentally
which has not been accomplished for many other circuits.

In numerical simulations, one often exploits the following dimensionless form of \Cc
\begin{equation}
\dot{\mathbf x}=
[\alpha(x_2-x_1-g(x_1)),
x_1-x_2+x_3,
-\beta x_2]^T
\label{Chua-norm}
\end{equation}
where $g(x_1) =b x_1 + {1 \over 2} (a-b)[|x_1+1|-|x_1-1|]$.
For the following parameter values $a = -1.27$, $b = -0.68$, $\alpha = 10.0$, $\beta = 14.87$
\Cc\ exhibits chaotic behaviour.

\begin{figure}[ht]
\begin{center}
\includegraphics[width=70mm, height=43mm]{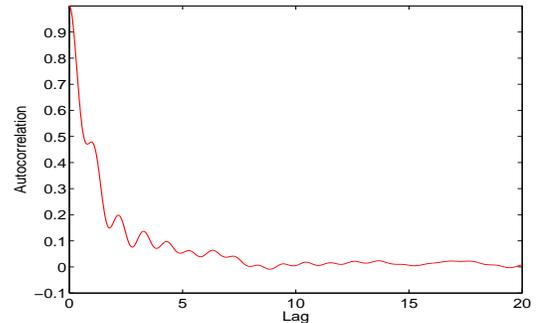}
\end{center}
\vspace{-6mm}
\caption{Autocorrelation of the chaotic signal generated by Chua's circuit.} 
\label{fig-ChuaCorr}
\end{figure}

The chaotic attractor of \Cc\ for the previous parameter values is widely known
as the {\em double-scroll\/} chaotic attractor.
As an indication of the randomness of chaotic trajectories 
we show in Fig.~\ref{fig-ChuaCorr} the normalised autocorrelation 
$$R_{x_1,\mbox{norm}}(\tau) = R_{x_1}(\tau)/R_{x_1}(0)$$
of the signal $x_1(t)$ generated by \Cc\ (\ref{Chua-norm}),
where 
$$R_{x_1}(\tau ) = \lim_{T \to \infty} {1 \over T}\int_{-T/2}^{T/2} x_1(t+\tau )x_1(t)\mbox{d}t.$$
Figure~\ref{fig-ChuaCorr} reveals rapid decorrelation between close samples, 
thus making this sequence akin to a random trajectory.

\vspace{-3mm}

\begin{figure}[ht]
\begin{center}
\includegraphics[width=70mm, height=43mm]{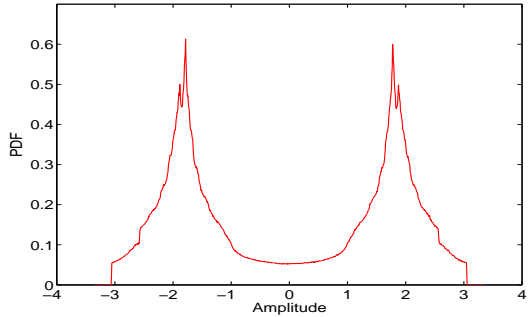}
\end{center}
\vspace{-6mm}
\caption{Probability density function of the chaotic signal generated by Chua's circuit.} 
\label{fig-ChuaPdf}
\end{figure}

Figure~\ref{fig-ChuaPdf} shows the probability density function of the chaotic signal $x_1(t)$ generated by \Cc\ (\ref{Chua-norm}).

%
%

\subsection{Lorenz} \label{sec:Lorenz}
We have numerically examined the Lorenz system
\begin{equation}
\dot {\mathbf x} = [16.0(x_2-x_1),45.6x_1-x_1x_3 - x_2, x_1x_2-4.0x_3]^T.
\label{eq-Lorenz}
\end{equation}

\begin{figure}[ht]
\begin{center}
\includegraphics[width=70mm, height=43mm]{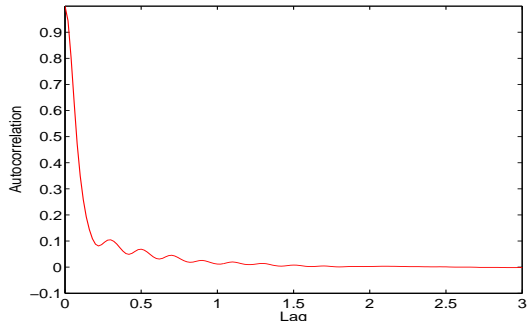}
\end{center}
\vspace{-6mm}
\caption{Autocorrelation of the chaotic signal generated by Lorenz system.} 
\label{fig-LorenzCorr}
\end{figure}

Correlation between samples decreases rapidly with the sampling distance, as shown in Figure~\ref{fig-LorenzCorr}.

\begin{figure}[ht]
\begin{center}
\includegraphics[width=70mm, height=43mm]{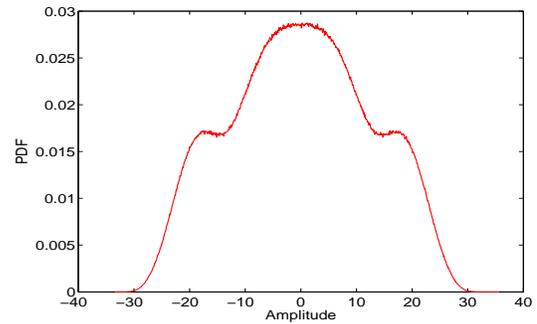}
\end{center}
\vspace{-6mm}
\caption{Probability density function of the chaotic signal generated by Lorenz system.} 
\label{fig-LorenzPdf}
\end{figure}

Figure~\ref{fig-LorenzPdf} shows the probability density function of the chaotic signal $x_1(t)$ generated by Lorenz system (\ref{eq-Lorenz}).

\vspace{-3mm}

\subsection{R\"ossler} \label{sec:Rossler}
We have also analysed the R\"ossler system 
\begin{equation}
\dot {\mathbf x} = [-x_2-x_3,x_1+0.2x_2,0.2+x_3(x_1-5.7)]^T.
\label{eq-Rossler}
\end{equation}

However, R\"ossler system spends a significant portion of the time making loops in the $x_1-x_2$ plane around the origin and spiralling outwards.
As $x_1$ approaches value 5.7, $x_3$ starts to climb.
This growth in $x_3$ makes $x_1$ to drop and the spiraling motion in the $x_1-x_2$ plane starts again.
This long spiraling motion causes strong periodic correlation in $x_1(t) $.
Consequently, unless very long sampling period is employed, generated sequences of numbers will be statisticaly dependent and the resulting chaos filters will not be useful for signal acquisition and reconstruction in compressed sensing.
This indicates that due attention needs to be paid to the choice of the chaotic dynamical system used in compressive sampling.

\section{Main results} \label{sec:MainResults}

In this section we explain how we use chaotic signals to sample 
sparse signals. We compare their performance to the performance 
when sampling is done by random measurement matrices. We also investigate 
the impact of correlation in the measurement sequence, chaotic or random.  

As an input sequence we use time-sparse signals:  $k$ spikes $\in\{-1, +1\}$ randomly 
set in a sequence of $N$ samples, with equal probability of $-1$'s and $1$'s.
A Bernoulli sequence of total of $k$ $-1$'s and
$1$'s with probability 0.5 is generated by using Monte Carlo simulation. 
Then $N-k$ zeros are added at the end of this sequence. 
Finally, the obtained sequence of length $N$ is randomly permuted.

Next we explain the process of construction of the measurement matrix $\Phi\in R^{M\times N}$. 
The procedure is the same for both random signals and chaotic signals.
We first generate a sequence ${\bf c}=[c_0, c_1, \ldots ,c_{MN-1}]$ of length $M\times N$ 
and then we construct a matrix 
of size $M\times N$ columnwise (taking $M$ contiguous samples and putting them in
a column of $\Phi$).

All elements of the matrix are scaled with the factor $1/(\sigma \sqrt{M})$ where $\sigma^2$ is the variance of the sampled signal (or sequence) used to construct the 
measurement matrix.
Thus
\begin{equation}
  \Phi =\frac{1}{\sigma \sqrt{M}}\left[      
   \begin{array}{cccc}
     c_0   &  c_{M} & \cdots & c_{M(N-1)} \\
     c_1   &  c_{M+1} & \cdots & c_{M(N-1)+1}  \\
     \vdots & \vdots & \vdots & \vdots \\
     c_{M-1}  &  c_{2M-1}  & \cdots & c_{MN-1} \\
   \end{array}
   \right]
\end{equation}
The vector of measurements $  {\bf y}\in R^M$ is obtained as
\begin{equation}
  {\bf y}=\Phi {\bf x}
\end{equation}

The problem of reconstruction can be stated formally as:
\begin{equation}
  \min _{{\bf s}\in R^N}\|{\bf s}\|_1  \mbox{  subject to  } {\bf y}=\Theta {\bf s}
\end{equation}

This is a well-established problem known as basis pursuit. Basis pursuit problems can be easily  transformed into a linear programming problem.



In our simulations we used the primal-dual interior point algorithm, whose Matlab 
implementation can be found here: 
http://www-stat.stanford.edu/$\sim$candes/software.html 
under L1 MAGIC.
It requires as an input the initial guess ${\bf x}_0$ for the solution, the
measurement matrix $\Phi$,  
the measurements ${\bf y}$, and the precision to which we want the problem solved. 

We performed extensive simulations with various measurement matrices.
We used two continuous time \CDS s for creating the elements of the measurement matrix: \Cc\ and Lorenz system.
For \Cc\  we sampled $x_1$ at sampling distance $\tau = 1$, while for Lorenz system we sampled $x_1$ at sampling distance $\tau = 0.5$.

We also used several random measurement matrices, obtained from sequences 
such as iid Gaussian, Gaussian with autocorrelation $R(i)=\rho ^{|i|}$, $i\in Z$ (we used $\rho =0.99$), Bernoulli, iid uniform in the range $[0, 1]$ and iid uniform in the range $[-0.5, 0.5]$.

\begin{figure}[ht]
\begin{center}
\includegraphics[width=80mm, height=50mm]{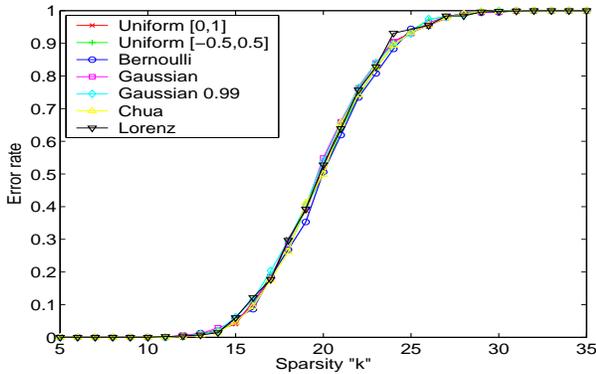}
\end{center}
\vspace{-6mm}
\caption{Probability of incorrect reconstruction $\Pr\{ e= ||\mathbf x-\mathbf x_r||/\|\mathbf x\|>\varepsilon =0.01\}$ as a function of the signal sparsity $k$ for $N=100$ and $M=50$.}
\label{fig-ErrorRate}
\end{figure}

Figure~\ref{fig-ErrorRate} depicts the dependence of the probability of 
incorrect reconstruction 
on the signal sparsity $k$ when 
$N=100$ and $M=50$. 
As a criterion for exact reconstruction we used that the relative error $e$ is smaller than $\varepsilon = 0.01$
\begin{equation}
e= ||\mathbf x-\mathbf x_r||/\|\mathbf x\|<\varepsilon =0.01
\end{equation}
where ${\bf x}_r$ is the reconstructed vector and $\|\cdot\|$ is the $l_2$ norm. 

As depicted in Fig.~\ref{fig-ErrorRate}, there are no significant differences in the performance of diferent measurement matrices. 
This is despite the significant normalised autocorrelation $R_{x_1,\mbox{norm}}(\tau=1) = 0.47$ for \Cc\ (\ref{Chua-norm}). 

Gaussian sequence with correlation $0.99$ also did not exhibit any noticeable performance
loss. We should keep in mind that since the elements of the measurement sequence
are written columnwise in matrix $\Phi$, the correlation is significantly decreased
(the adjacent elements in each row come from elements in the sequence that are $M$ samples apart).

\vspace{-3mm}
%
%

\begin{figure}[ht]
\begin{center}
\includegraphics[width=80mm, height=50mm]{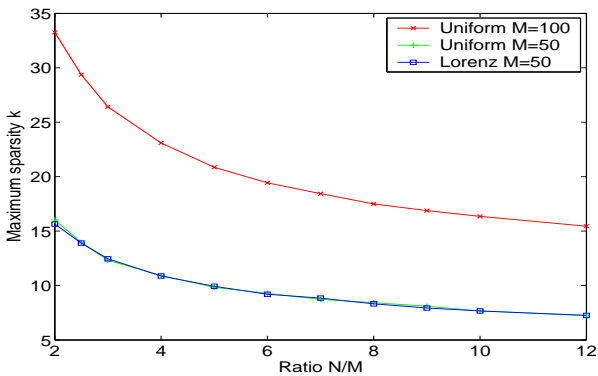}
\end{center}
\vspace{-6mm}
\caption{Dependence of $k_{max}$ on $N/M$.}
\label{fig-kOdmn}
\end{figure}

Maximum sparsity $k_{max}$ which allows for correct recovery of ${\bf x}$ does not depend on the measurement matrix, and is solely determined by $N$ and $M$.
Figure \ref{fig-kOdmn} depicts the dependence of the maximum sparsity $k_{max}$ on the ratio $r=N/M$ for $\varepsilon =0.01$.
$k_{max}$ is determined as the maximum sparsity for which error rate is smaller than $0.1$. Linear interpolation is used to determine $k_{max}$.
If $M$ is doubled, then $k_{max}$ increases roughly by factor $2.1$ for each $N$.

%
%

\begin{figure}[ht]
\vspace{-3mm}
\begin{center}
$\begin{array}{cc}
\hspace{-5mm}
\includegraphics[width=47mm, height=37mm]{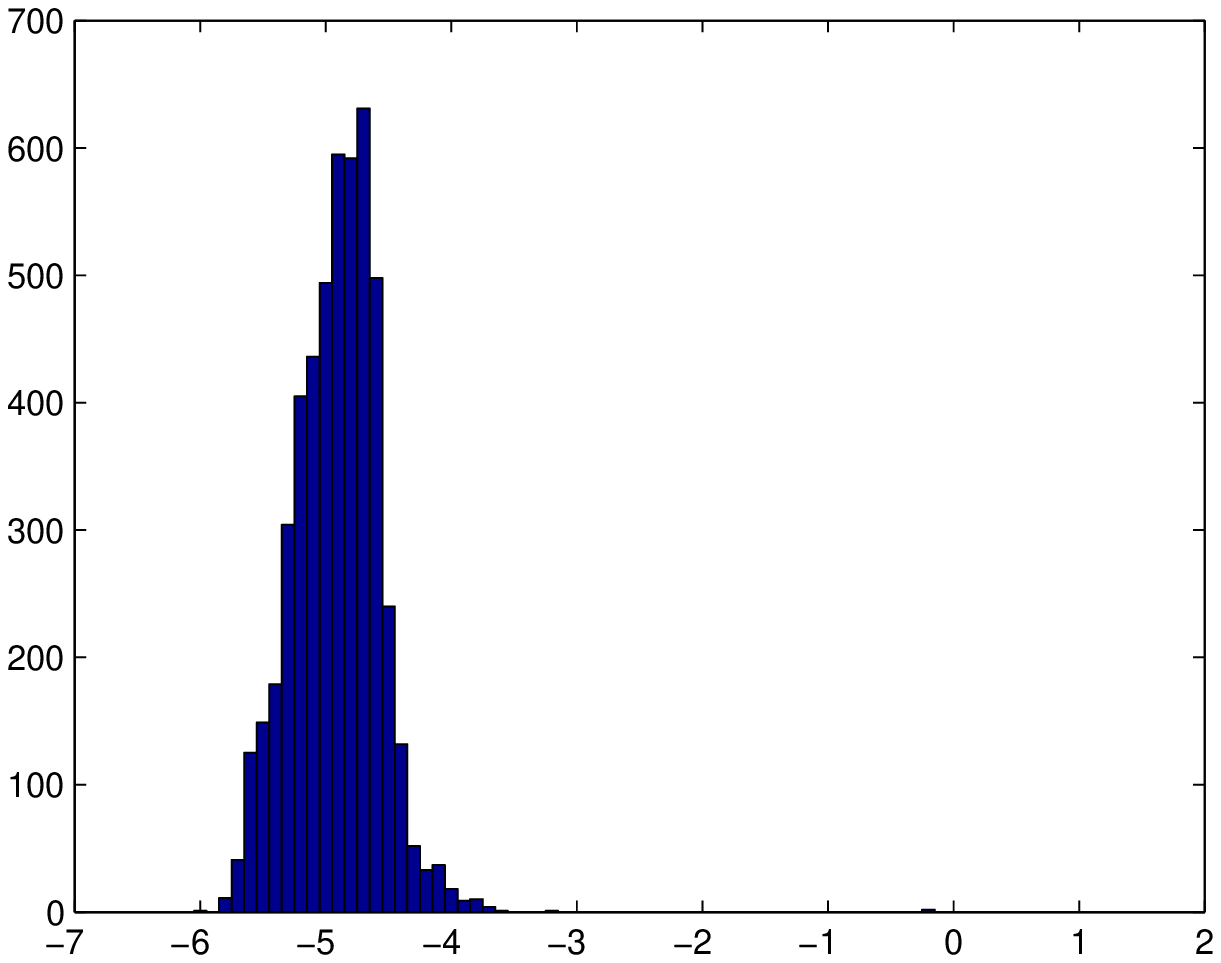} &
\includegraphics[width=47mm, height=37mm]{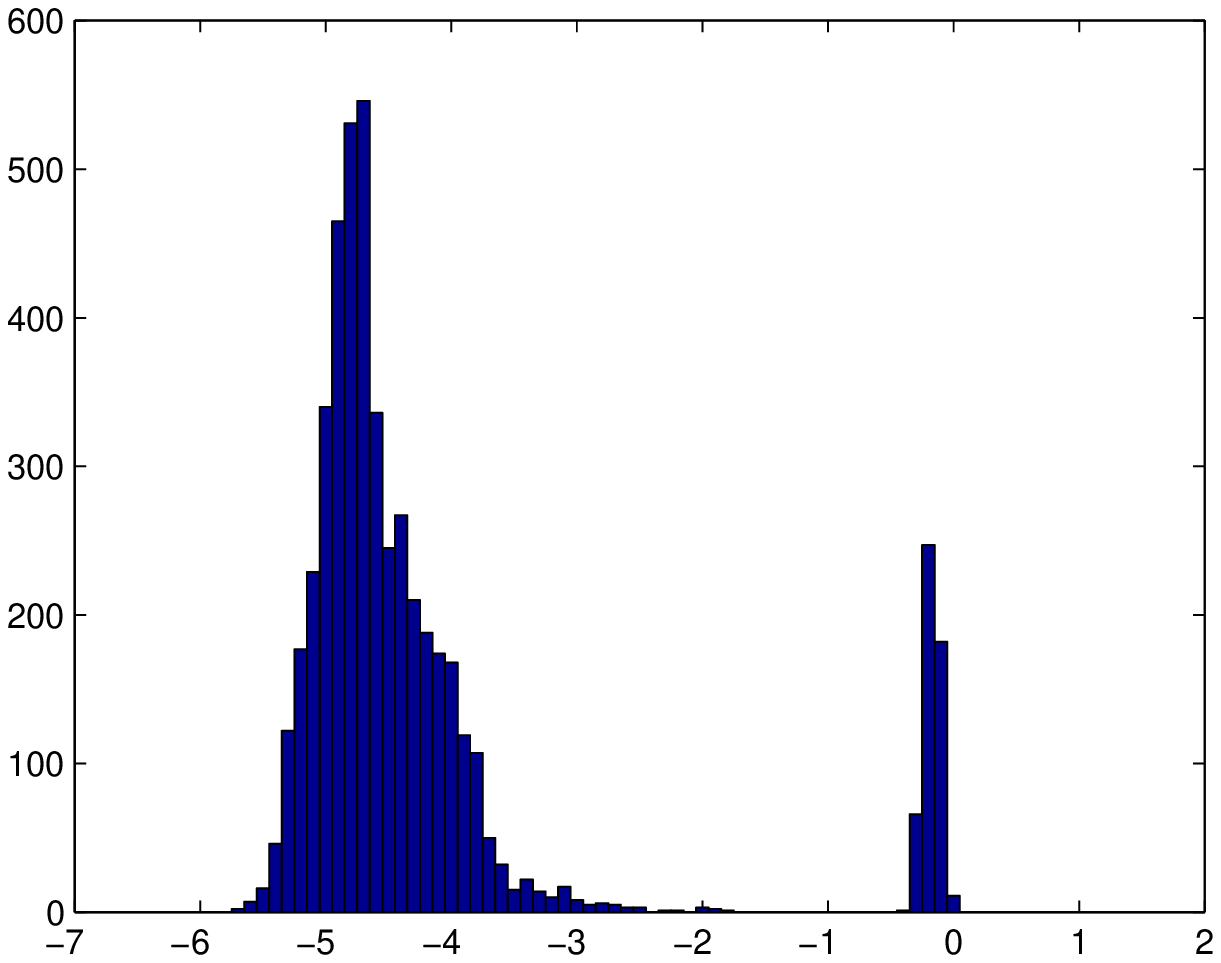} \\
(a) & (b) \\
\hspace{-5mm}
\includegraphics[width=47mm, height=37mm]{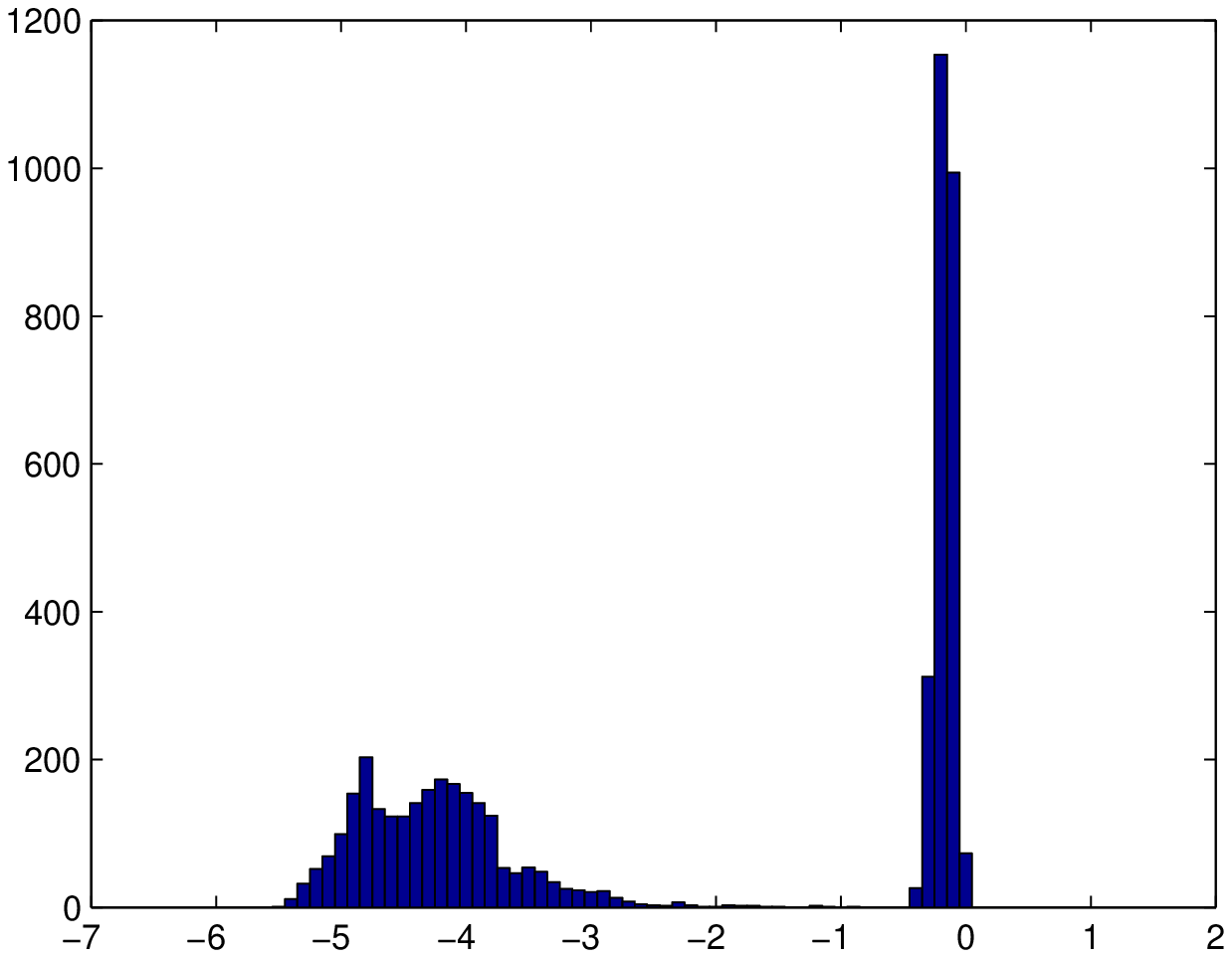} &
\includegraphics[width=47mm, height=37mm]{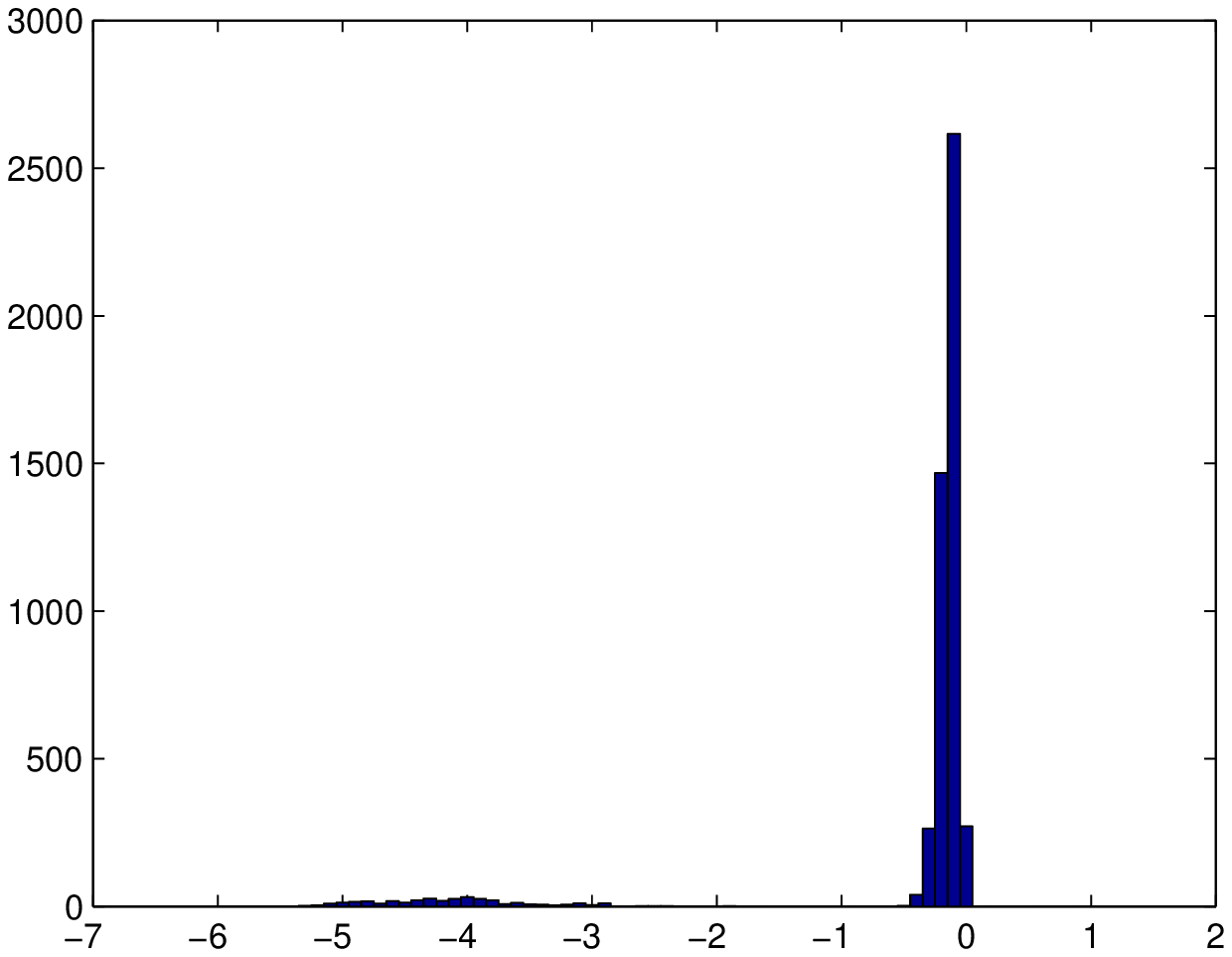}  \\
(c) & (d) 
\end{array}
$
\end{center}
\vspace{-3mm}
\caption{Histograms of the logarithm of the relative error $e=||\mathbf x-\mathbf x_r||/||\mathbf x||$. (a) k=11; (b) k=16; (c) k=20; (d) k=25}
\label{fig-ErrorPdf}
\end{figure}

Figure~\ref{fig-ErrorPdf} depicts histograms of the logarithm of the relative error between original and recovered signal $e=||\mathbf x-\mathbf x_r||/||\mathbf x||$ for fixed $N=100$ and $M=50$, and for various $k$, obtained from $5,000$ simulation runs. 
When the original signal is correctly recovered, then a small but finite error $e$ occurs due to the finite precision to which the optimization problem is solved by the L1 MAGIC algorithm.
If the recovery is incorrect, then a large relative error occurs. Consequently, wide range of values of $\varepsilon\in[10^{-2.5}, 10^{-0.5}]$ can distinguish between the correct and incorrect recovery.

\vspace{-5mm}

\section{Conclusion} \label{sect:conclusion}

We studied the use of different chaotic signals to 
construct measurement matrices in compressive sampling.
We showed that Chua and Lorenz chaotic signals
show performance comparable to that of 
random Gaussian, Bernoulli and uniformly distributed sequences. 
We determined that the correlation does not increase the 
probability of incorrect reconstruction. The performance is relatively 
insensitive on the value of the parameter $\varepsilon$ used
to test the exact reconstruction. 

\ifCLASSOPTIONcaptionsoff
  \newpage
\fi
\bibliographystyle{IEEEtran}
\bibliography{ref_CS_new}
\end{document}